\documentclass[acmtog,balance=false]{acmart}
\AtBeginDocument{%
  }

\setcopyright{acmlicensed}
\copyrightyear{2025}
\acmYear{2025}
\acmYear{2025} 
\setcopyright{acmlicensed}\acmConference[SIGGRAPH Conference Papers '25]{Special
Interest Group on Computer Graphics and Interactive Techniques Conference
Conference Papers }{August 10--14, 2025}{Vancouver, BC, Canada}
\acmBooktitle{Special Interest Group on Computer Graphics and Interactive
Techniques Conference Conference Papers (SIGGRAPH Conference Papers '25), August
10--14, 2025, Vancouver, BC, Canada}
\acmDOI{10.1145/3721238.3730705}
\acmISBN{979-8-4007-1540-2/2025/08}

\acmSubmissionID{828}

\usepackage{tikz}
\usepackage{xspace}
\usepackage{subcaption}
\usepackage{xcolor}
\usepackage{float}
\usepackage{placeins}

\definecolor{revision}{rgb}{0.0, 0.0, 0.0}

\makeatletter
\DeclareRobustCommand\onedot{\futurelet\@let@token\@onedot}
\def\@onedot{\ifx\@let@token.\else.\null\fi\xspace}

\def\eg{\emph{e.g}\onedot}

\def\etc{\emph{etc}\onedot}

\makeatother

\newcommand{\ourmethod}{\emph{BuildingBlock}\xspace}

\begin{document}

\title{\ourmethod: A Hybrid Approach for Structured Building Generation}

\author{Junming Huang}
\authornote{Equal contributions.}
\email{junminghuang@zju.edu.cn}
\orcid{0009-0001-5284-9431}
\author{Chi Wang}
\authornotemark[1]
\email{wangchi1995@zju.edu.cn}
\affiliation{%
  \institution{Zhejiang University}
  \city{HangZhou}
  \country{China}
}

\author{Letian Li}
\email{3210103854@zju.edu.cn}
\affiliation{%
  \institution{Zhejiang University}
  \city{HangZhou}
  \country{China}
}

\author{Changxin Huang}
\email{changxinhuang@zju.edu.cn}
\affiliation{%
  \institution{Zhejiang University}
  \city{HangZhou}
  \country{China}
}

\author{Qiang Dai}
\email{jondai@lightspeed-studios.com}
\authornote{Corresponding authors.}
\affiliation{%
  \institution{LIGHTSPEED}
  \city{HangZhou}
  \country{China}
}
\author{Weiwei Xu}
\email{xww@cad.zju.edu.cn}
\authornotemark[2]
\affiliation{%
  \institution{State Key Lab CAD\&CG, Zhejiang University}
  \city{HangZhou}
  \country{China}
}

\renewcommand{\shortauthors}{Junming Huang, Chi Wang, Letian Li, Changxin Huang, Qiang Dai and Weiwei Xu}

\begin{abstract}
  Three-dimensional building generation is vital for applications in gaming, virtual reality, and digital twins, yet current methods face challenges in producing diverse, structured, and hierarchically coherent buildings. We propose \ourmethod, a hybrid approach that integrates generative models, procedural content generation (PCG), and large language models (LLMs) to address these limitations. Specifically, our method introduces a two-phase pipeline: the Layout Generation Phase (LGP) and the Building Construction Phase (BCP). 
  LGP reframes box-based layout generation as a point-cloud generation task, utilizing a newly constructed architectural dataset and a Transformer-based diffusion model to create globally consistent layouts. With LLMs, these layouts are extended into rule-based hierarchical designs, seamlessly incorporating component styles and spatial structures. 
  The BCP leverages these layouts to guide PCG, enabling local-customizable, high-quality structured building generation. Experimental results demonstrate \ourmethod’s effectiveness in generating diverse and hierarchically structured buildings, achieving state-of-the-art results on multiple benchmarks, and paving the way for scalable and intuitive architectural workflows.
\end{abstract}


\begin{teaserfigure}
  \includegraphics[width=\textwidth]{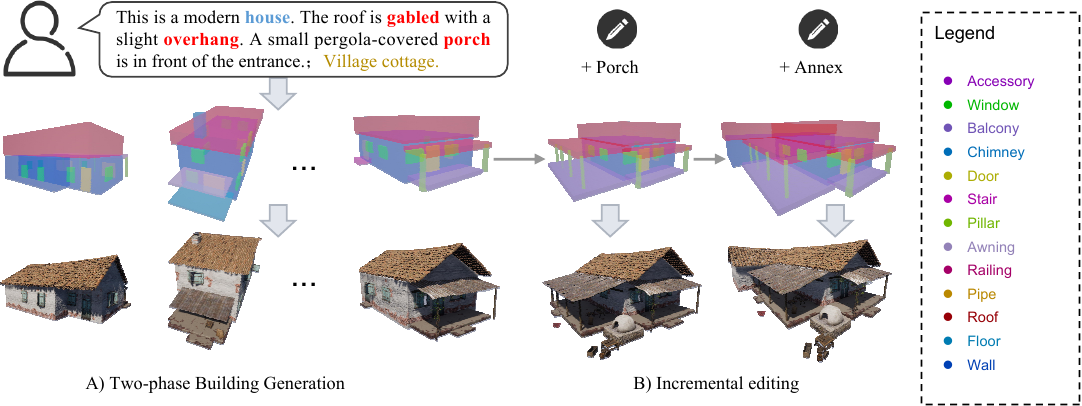}
  \caption{
  Our method can faithfully generate structured buildings from text prompts. 
  From top to bottom: user instruction, box-based layout, structured building.
  A) The same text prompt can generate diverse layouts, which can further generate diverse buildings.
  B) An example of incremental local editing for an instance of part A. 
  It can be observed that our method can quickly generate high-quality results according to the local edits of a box-based layout, while keeping the others unaffected.
  }
  \label{fig:teaser}
\end{teaserfigure}


\maketitle

\section{Introduction}
\label{sec:intro}

\newcommand{\TODO}[1]{{\color{red}{TODO:#1}}}

The modeling of three-dimensional (3D) building generation plays an important role in computer graphics, powering applications in gaming, virtual reality, architecture, and digital twins.
The procedural content generation~(PCG)~\cite{nishida2016interactive,hu2021ChineseAncientPCG} has garnered significant attention, given its ability to automate the design of buildings. However, they still face the challenge of generating diverse buildings that meet the high-level descriptions.

The current generative models~\cite{zhang2024clay, meshy, li2024instant3d, sun2023dreamcraft3d, zhou2024dreampropeller, yi2024gaussiandreamer} pave a new way for the diverse text-driven building generation. However, those generated contents lack structured hierarchical information, such as floor count and intricate details. 
While it is possible to train a model for structured hierarchical building generation, it is challenging for structured dataset collection and model architecture design. 



Inspired by pioneers~\cite{10.1145/2185520.2185551, talton2011metropolis} employing the probabilistic model as a procedural approach guide, we propose a novel hybrid approach \ourmethod to address these limitations. 
\ourmethod has a two-phase strcuture, that integrates a generative model and LLMs with PCG to the structured hierarchical building generation with highly flexible user descriptions.
This hybrid approach allows for global control of PCG via controlling the placement of building components (like roof, wall, and door) which are composed of a group of ``LEGO blocks''.
In detail, \ourmethod consists of Layout Generation Phase~(LGP) and Building Construction Phase~(BCP). 
LGP first utilizes a generative model to produce box-based layouts, which are then extended to rule-based layouts with the help of LLMs. 
With the rule-based layout, BCP leverages PCG to generate structured buildings. This hybrid cooperation ensures that structural components are both locally detailed and globally coordinated. 
Moreover, this hybrid design scheme is inherently generalizable, offering a framework that can be adapted to other controllable structured generative tasks, where the seamless coexistence of global control, high-level constraints, and localized adjustments is essential.

Existing works~\cite{paschalidou2021atiss,tang2024diffuscene} excel at producing plausible box-based 3D indoor layouts and are widely adopted. 
However, they cannot be directly applied in LGP for buildings due to the following two reasons. 
1) imperfect match between the unordered data and CNN architectures or autoregressive structures. Although efforts like z-order sorting have been made to reduce the sequence order impact, there is still space for improvement. We employ a Transformer-based~\cite{vaswani2017attention} diffusion~\cite{rombach2022stable_diffusion, sohl2015deep} network without positional encoding to completely eliminate the order dependency, ensuring that rearranging the placement order of components in the sequence does not affect the overall semantic meaning.
2) lack of available architectural layout dataset.
To bridge this gap, we construct a novel architectural layout dataset, with paired data of box-based layouts and corresponding descriptions. 
Building upon this dataset, we are able to train layout generation networks to create layouts with global consistency. 

The box-based layout generation in LGP is treated as a point cloud generation task, considering it from the perspective of unordered sequences. We design a Point$\cdot$E~\cite{nichol2022pointegenerating3dpoint}-based network to address the specific challenges of layout generation, enabling the production of high-quality layouts. 
Furthermore, a key insight is that box-based layouts can be further extended into rule-based layouts with the assistance of large language models~(LLMs). This extension enables the incorporation of component styles and hierarchical structures into the spatial layouts, facilitating structured expression at the layout level without compromising diversity. 
The contributions of this work are summarized as follows:
\begin{itemize}
    \item We introduce a novel hybrid approach for text-driven structured building generation via a combination of the traditional PCG, generative models, and LLMs. A structured rule-based layout is generated with the generative model and LLMs, which is then fed into PCG to produce the structured hierarchical buildings.
    \item We present a novel Transformer-based diffusion model to generate box-based building layouts, as a global structured guidance for further building generation. 
    With the help of LLMs, these layouts are further extended into rule-based designs, integrating component styles and hierarchical structures.
    \item We propose a 3D architectural layout dataset which contains around 1.2k buildings, and more than 40k blocks, along with 9.6k corresponding rendered images and descriptions from eight different viewpoints. It can be used for layout generation, architectural component detection, and other related fields.
\end{itemize}

\ourmethod can generate high-quality diverse structured buildings while also enabling user-friendly editing of layouts, allowing for high-quality and globally consistent building modifications without altering other parts (Fig.~\ref{fig:teaser}).
Experimental results demonstrate the superiority of our method in generating plausible architectural layouts in challenging scenarios, achieving substantial improvements over other methods across all quantitative metrics.
In addition, we further evaluate \ourmethod on the 3D-FRONT indoor dataset, achieving state-of-the-art results, underscoring the robustness and applicability of our approach.

\section{Related Work}
\begin{figure*}[t]
  \includegraphics[width=\textwidth]{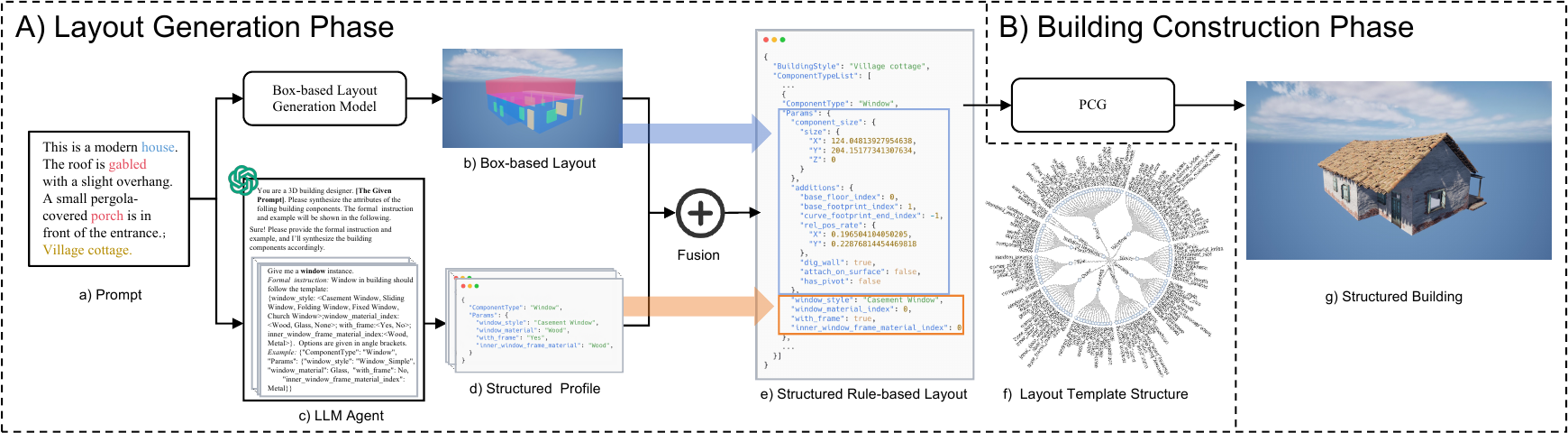}
  \caption{The pipeline of \ourmethod. Given a text description for the building, a stylized layout profile is first generated, which is obtained by fusing two phases. A) The \emph{Layout Generation Phase} has a dual-branch structure, one diffusion-based branch produces bounding box generation for the layout, and the other LLM-based branch provides the style information extractor for the components. B) Then, the \emph{Building Construction Phase} employs a PCG method to generate the corresponding structured building based on the stylized layout profile. Best view on screen and zoom in.}
  \Description{}
  \label{fig:pipeline}
\end{figure*}

\paragraph{Procedural Modeling}
Procedural content generation~\cite{shaker2016procedural, haglund2009procedural, togelius2011search, yannakakis2011experience} is a set of methods that automatically create digital content with predefined rules, such as geometric fractals and behavioral simulations.
Early works focused on natural phenomena such as terrains~\cite{perlin1985image} and vegetation~\cite{aristed1968mathematical}, using fractals and noise functions to simulate randomness. 
Perlin noise~\cite{perlin1985image, lagae2010survey} is a gradient-based procedural noise technique that interpolates smoothly between pseudo-random gradient vectors on a grid, enabling efficient generation of continuous, natural-looking patterns for textures, terrains, and other stochastic content.
As an extension, grammar-based methods~\cite{santoni2016gtangle, wonka2003instant, karnick2009shape, larive2006wall}, such as Graph Grammar~\cite{christiansen2012generic} and Generative Modeling Language~\cite{haglund2009procedural}, became prominent due to their ability to express complex combinations of rules. 
Grammar-based methods play an important role in difficult tasks like building modeling~\cite{adao2014procedural, muller2006procedural, nishida2018procedural} and 
city creation~\cite{parish2001procedural, kelly2007citygen, kim2018procedural}, offering both automation and user customization, with applications in city planning, road placement, and building generation. 
Pascal M\"uller~\cite{muller2006procedural} utilizes a shape grammar with pre-defined rules to achieve the building design by iteratively adding details.
These methods enable the efficient generation of large-scale fine-detail scenes, automate repetitive tasks, and allow flexible adaptation to various styles and levels of detail. However, they often lack global control over the overall structure, making it difficult to ensure coherence among components.

\paragraph{Deep-learning-based generation}
Deep-learning-based generation leverages advanced neural networks~\cite{xu2023survey}, such as generative adversarial networks~\cite{goodfellow2020GAN, karras2019style, donahue2019large} and diffusion models~\cite{rombach2022stable_diffusion, ho2020denoising, yang2023diffusion}, to create digital content with a focus on component coordination and global structural control. 
Unlike traditional PCG methods, deep learning excels at learning complex patterns and relationships from data, ensuring that the generated content is both coherent and realistic. 
Recent deep-leaning-based scene generation can be grouped into two primary categories: single-stage and two-stage generation. 
In the single-stage generation, a neural network directly produces a complete scene in an end-to-end manner with implicit~\cite{zhang20243d, zhang2024text2nerf} or explicit~\cite{meng2024lt3sd} scene representation.
These approaches are efficient and ensure global coherence, but offer limited flexibility for modifying specific components and hard to achieve the precise object-to-object alignment of PCG.
To address this issue, two-stage generation, on the other hand, separates the process into layout generation and content population~\cite{lin2024genusd, bahmani2023cc3d, po2024compositional}. In the layout generation stage, a model generates a scene layout (e.g., classes, sizes, and positions of objects). 
In the content population stage, generated, retrieved, or predefined objects are populated with the layout, allowing for more customization while maintaining the overall structure. 
LEGO~\cite{wei2023lego} views the scene generation problem as a rearrangement task for a given initial scene, aiming to obtain the desired scene by optimizing the layout state.
Deep-learning-based scene generation methods, while addressing the issue of global consistency in procedural methods, introduce a new problem: the generated scenes often lack structured hierarchical representations, making it more challenging for designers to select and edit components.

\paragraph{LLM Integration.} In recent years, with the rise of large language models~(LLMs), their powerful reasoning and generalization capabilities have garnered significant attention. LLMs have successfully served as decision-makers in various downstream tasks, such as code generation~\cite{zhu2024deepseek, nijkamp2023codegen2}, and understanding of autonomous driving scenes~\cite{yang2023survey, feng2024layoutgpt}.
\textcolor{revision}{LayoutGPT~\cite{feng2024layoutgpt} employs LLMs for layout-based visual planning in multiple domains. However, this zero-shot approach relying solely on LLMs struggles to effectively capture the complex structures in building layout.}
SceneX~\cite{zhou2024scenex} introduces the LLM agent into PCG software such as Blender to provide automatic 3D asset generation.
Similarly, our method uses LLMs to better understand user inputs. Unlike \textcolor{revision}{SceneX}, which directly employs PCG for content generation, our method combines knowledge from LLMs with AIGC-generated layouts. These are then incorporated into a structured rule-based layout to guide PCG. This design leverages generative models for the specific task, while also enabling general LLM’s reasoning capabilities and allowing users to easily edit the intermediate structured layout.

DiffuScene~\cite{tang2024diffuscene} is the most closely related method to \ourmethod, as it synthesizes scene layouts using a diffusion model, and then queries the object database for the closest one. Similarly, \ourmethod also generates layouts in the first stage, but in the content population stage, we leverage a PCG approach to create more detailed and adaptive scenes with structured hierarchical representations based on the layouts. 

\section{Building Blocks}

In this chapter, we introduce \ourmethod, a text-driven automated pipeline for structured building generation~(Fig.~\ref{fig:pipeline}). The pipeline consists of two phases: Layout Generation Phase (LGP) and Building Construction Phase (BCP). The LGP generates stylized layouts from the given building description. A Transformer-based diffusion model is used to produce attributes such as component positions, dimensions, and categories, while a large language model (LLM) generates stylized information for the components, including materials, internal structures, \etc~(Fig.~\ref{fig:pipeline}A). 
The BCP employs procedural content generation (PCG) methods to refine and construct structured buildings based on the given layout~(Fig.~\ref{fig:pipeline}B).

To formalize the process, we assume that each building consists of no more than $N$ component boxes $\{C_i\}_{i=1}^N$, represented as an unordered set. 
\textcolor{revision}{Each component in this set is denoted as $C_i=[l_i,s_i,c_i]$, where $l_i \in \mathbb{R}^{3}$, $s_i \in \mathbb{R}^{3}$, and $c_i \in \mathbb{R}$ is it's position, size, and class attribute, respectively. Since the number of boxes varies across buildings, following the approach used in DiffuScene~\cite{tang2024diffuscene}, we extend the class attribute by introducing an empty class.
}
Empty-class boxes are used to pad the total number of boxes in each building to the fixed value $N$.  
Unlike DiffuScene~\cite{tang2024diffuscene}, we randomly assign other attributes (\eg, position, size) to the empty-class boxes based on the non-empty ones, denoted as \emph{PadReal}. 
Experimental results demonstrate that this operation improves the stability of the network during training.
The box-based layout can be treated as an assembly of unordered boxes, where rearranging the boxes within a layout does not affect the overall semantic meaning. The sizes of buildings are normalized to reside in a global coordinate system, with the center of its bounding box located at $[0.5, 0.5, 0.5]$. 

In the following, we first provide a brief overview of the foundational knowledge regarding the diffusion model in the Subsec.~\ref{subsec:diffusion}. Next, we describe the construction of the dataset used in this paper in Subsec.~\ref{subsec:data_construction}. Finally, we explain the two phases, LGP and BCP, in Subsecs.~\ref{subsec:phase1} and ~\ref{subsec:phase2}, respectively.

\subsection{Preliminary}
\label{subsec:diffusion}
The DiffuScene~\cite{tang2024diffuscene} focuses on diverse and realistic 3D layout synthesis. It leverages a U-Net-based diffusion model to generate a box-based layout from textual prompts effectively. This model takes boxes as input and denoises them using MLP~\cite{taud2018multilayer} encoding, 1D-UNet, and MLP decoding.

Starting from random Gaussian noise, the model generates box-based layouts through a denoising process, which is optimized using a noise-based loss:
\begin{equation}
\mathcal{L}=\mathbb{E}_{C_0, \epsilon \sim \mathcal{N}(0, I), t}\left[\left\|\epsilon-\epsilon_\theta\left(C_t, t\right)\right\|_2^2\right],
\end{equation}
where $t$ represents the timestep, $C_0$ is the ground truth box-based layout, and $C_t$ is the noised layout. Noise scheduling is handled using a weighted combination of $C_0$ and Gaussian noise $\epsilon$, as $C_t = \sqrt{\bar{\alpha}_t} C_0 + \sqrt{1 - \bar{\alpha}_t} \epsilon$, with $\bar{\alpha}_t$ controlling the progression of noise during training.

In addition, the Intersection over Union~(IoU)~\cite{zhou2019iou} summation of arbitrary two bounding boxes is also supervised with an IoU loss:
\begin{equation}
\mathcal{L}_{iou}=\sum_{t=1}^N 0.1 * \bar{\alpha}_t * \sum_{C_i, C_j \in \tilde{\mathbf{x}}_0^t} \operatorname{IoU}\left(C_i, C_j\right),
\end{equation}
Where $\tilde{\mathbf{x}}_0^t$ is the approximation of a clean layout, which can be computed with Bayes’s theorem~\cite{ho2020DDPM}.

\subsection{Blocks Dataset Construction}
\label{subsec:data_construction}

We constructed a building dataset annotated with prompt descriptions, style labels, and bounding box-based layout, comprising 1.2k buildings and 42k boxes. The dataset was built from two sources: first, we extracted 1k buildings from the labeled dataset BuildingNet, converted the mesh annotation into bounding box annotations, and improved annotation quality through manual review and adjustments. Second, we collected additional building data from Sketchfab and Fab platforms and manually annotated their components with bounding boxes.
To ensure precise annotation of box positions and dimensions, we utilized Unreal Engine~(UE)~\cite{ue5, sanders2016introduction} and used the ``Cube'' shapes to label the bounding boxes of components. 
The categories and their distribution are shown in Fig.~\ref{fig:component_dist}.
The label for each building was made manually to ensure accuracy and consistency. 
As Fig.~\ref{fig:style_label} shows, our building label adopts a two-tier structure: first, they categorize building types, followed by optional attributes such as region and era.
To generate additional textual prompts corresponding to the 3D buildings, we employed multi-modal large language models through the following process: first, we rendered images of the buildings from 8 different viewpoints using UE. Then, we both utilized Gemini-Pro~\cite{team2023gemini} and ChatGPT-4~\cite{achiam2023gpt} to generate multi-view consistent textual descriptions for each building,
leveraging their complementary strengths to enhance prompt diversity.
Specifically, in the first step, descriptions are produced for each individual viewpoint, and in the second step, a globally consistent summary is synthesized based on multiview descriptions and corresponding images. Finally, all generated descriptions were manually reviewed, and redundant parts were removed, such as the infinity loop response. 
This process ensures high-quality annotation precision, style labels, and textual descriptions, providing a reliable foundation for subsequent research.

\subsection{Layout Generation Phase}
\label{subsec:phase1}

\begin{figure*}
    \centering
    \includegraphics[width=0.78\linewidth]{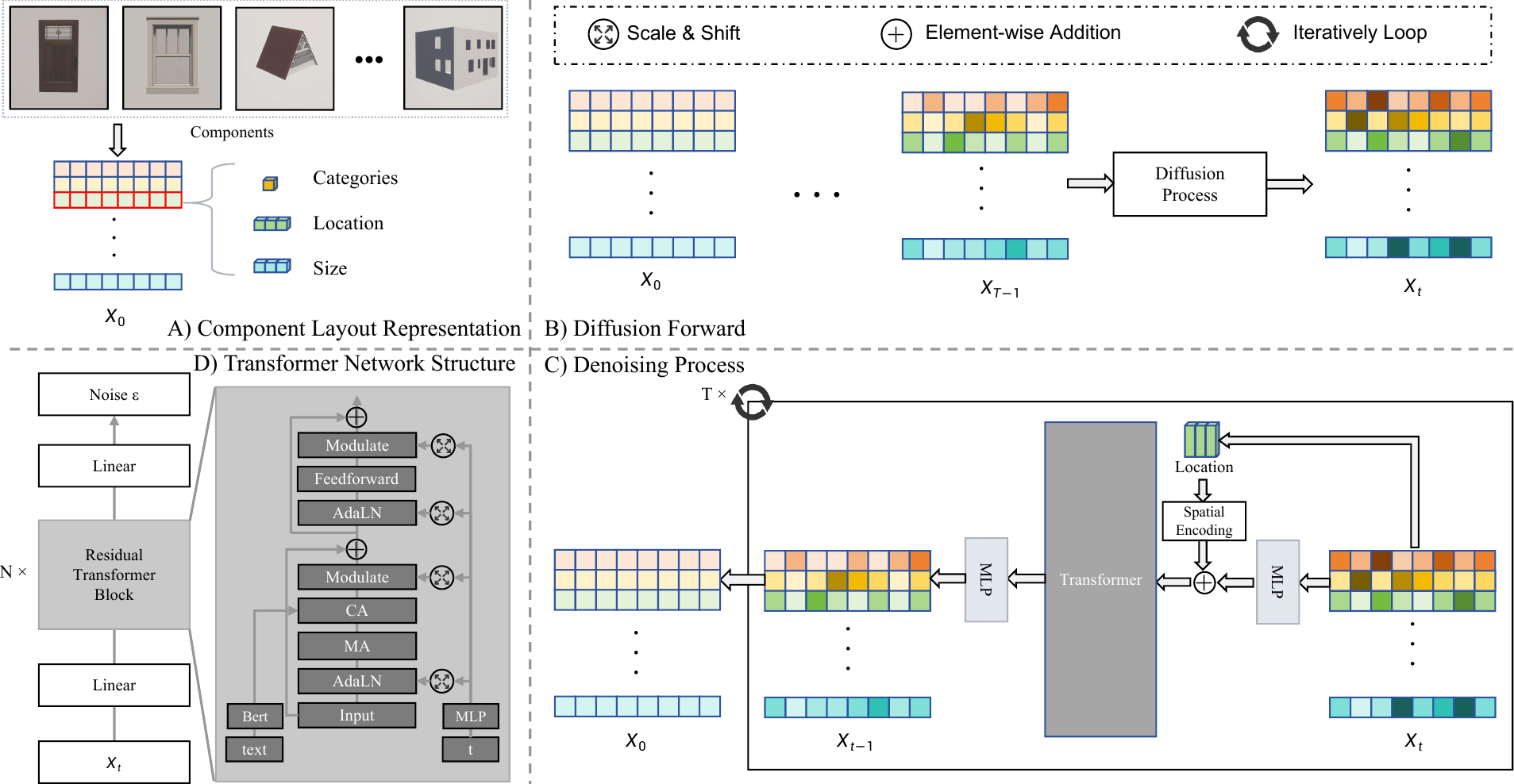}
    \caption{The process of box-based layout generation. A) The box-based layout representation, including categories, sizes, and locations. 
    B) The noise addition process of the layout diffusion model.
    C) The denoising process of the layout diffusion model, where the traditional positional encoding is removed and replaced by spatial encoding from the spatial positions of boxes.
    D) The Transformer network structure used in the diffusion model.
}
    \Description{}
    \label{fig:diffusion_model}
\end{figure*}

Layout generation is a core component of our method. Similar to traditional layout approaches, we first generate bounding box-based layouts. Building on this foundation, the key insight of this work is that box-based layouts can be further refined and enhanced into rule-based layouts. This can be achieved through the LLM agent’s ability to summarize and reason.

\paragraph{Box-based layout}

As Fig.~\ref{fig:diffusion_model} shows, our method adopts DiffuScene's pipeline~\cite{tang2024diffuscene} and replaces the original U-Net~\cite{ronneberger2015u} network with the Transformer in Point$\cdot$E network, which enables extensive interaction among unordered components.
To better leverage the combined capabilities of Transformers and diffusion models, we additionally adopt the adaptive layer normalization~(AdaLN) from DiT~\cite{Peebles2022DiT}. 
Meanwhile, a BERT~\cite{Devlin2019BERTPO}-based text control is integrated through a cross-attention mechanism. 
To get rid of the influence of the relative order in the Transformer input sequence, we reference Point$\cdot$E to remove the original positional encoding in the diffusion model. 
Instead, we introduce a spatial encoding, using a single-layer MLP structure to enhance the spatial information of box positions in the 3D space.
As shown in Fig.~\ref{fig:diffusion_model}C, the noised 3D locations are used to explicitly strengthen the spatial information of the input of the Transformer. 

\paragraph{Rule-based layout}
The rule-based layout expression has higher information capacity and flexibility than the bounding-box-based layout expression. 
It not only includes spatial and categorical information but also allows for further customization of attribute types and the representation of hierarchical structures. 
Moreover, a standardized expression can provide a foundation for subsequent PCG. 
We use the JSON format as the data interchange format of rule-based layouts, attempting to further extract and integrate user instructions. 
Specifically, we first generate the initial rule-based layout with a box-based layout~(Fig.~\ref{fig:pipeline}b) and layout templates for each type box~(categories in Fig.~\ref{fig:component_dist}). 
Fig.~\ref{fig:pipeline}\textcolor{revision}{(f)} visualizes the structural tree of layout templates. Due to the extensive number of key-value pairs in the full JSON, only the keys from the first two levels of the structure are displayed.
Then, we provide LLMs with building textual prompts~(Fig.~\ref{fig:pipeline}a) as a building prior. Based on the prior, it infers the style attributes for each box with a structured profile~(Fig.~\ref{fig:pipeline}c). For instance, LLMs can deduce that a modern office building might feature transparent or reflective glass windows with metal frames, even if not explicitly specified. 
These attributes overwrite the initial rule-based layout to enhance diversity, reasoning, and global consistency~(Fig.~\ref{fig:pipeline}e).

Rule-based layouts not only successfully accommodate the spatial and stylistic information of components but also can integrate the structural information of the layout. Our structured layout is centered around walls, and can determine the attachment relationships of components, such as doors and windows, based on relative positioning. This enables the construction of two-level structured layout information, providing prior hierarchical structure information for PCG. 
\textcolor{revision}{In addition, components can be further decomposed into sub-components to enable detailed adjustment.}

\subsection{Building Construction Phase}
\begin{figure}
    \centering
    \includegraphics[width=\linewidth]{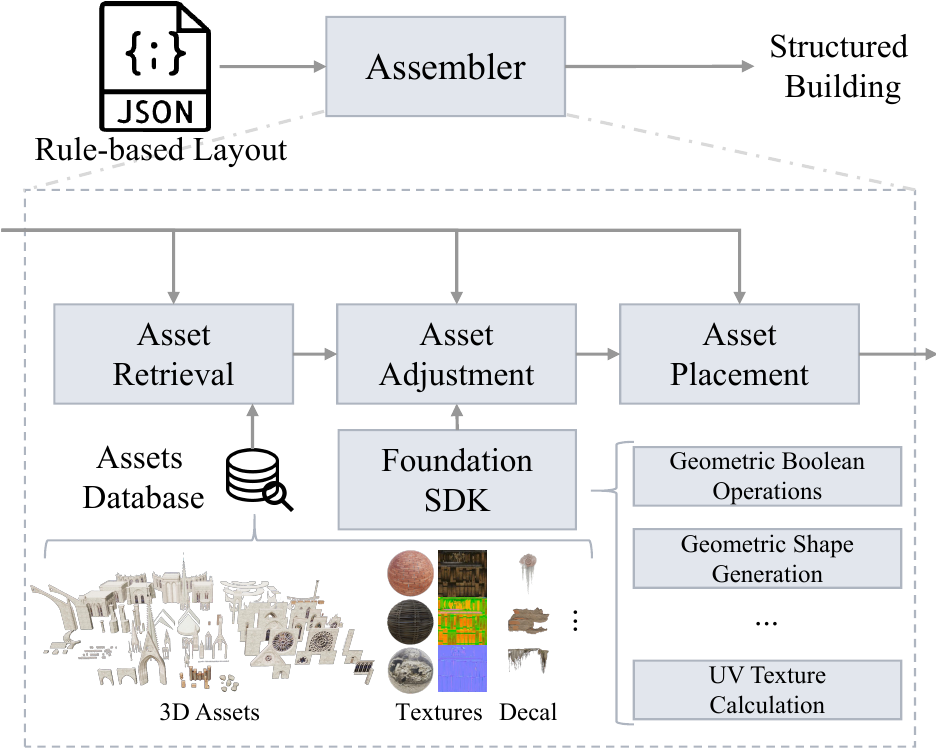}
    \caption{The pipeline of our PCG.}
    
    \Description{}
    \label{fig:phase2_pcg}
\end{figure}

In the Building Construction Phase, we apply the PCG method to achieve high-quality structured building generation with the rule-based layout. The pipeline of this phase is illustrated in Fig.~\ref{fig:phase2_pcg}.

First, for each required component in the rule-based layout, we retrieve matching components from the database \textcolor{revision}{ where each component possesses attributes such as category, style, and size ratio. After determining the component category, we first retrieve candidates by style, and then select the one whose size ratio most closely aligns with the desired value.}

Then, we handle detailed adjustments of the components by calling the Foundation SDK library, with each function implemented through Python functions. 
For example, Geometric Boolean Operations are employed to resolve various interactions between components, such as between windows and walls. In this case, the wall is first carved out to fit the window, and then the window can be inserted.
Besides, the overlapped walls are merged with CGAL~\cite{fabri2009cgal} efficiently.
Another example is the Geometric Scaling Function, which is used to handle the stretching of components. For the window case, a structured modern glass window consists of 
\textcolor{revision}{sub-components} like glass, the window frame, and the inner frame. The glass can be freely stretched or duplicated. However, directly stretching the window and inner frames may lead to uneven proportions. To achieve a reasonable stretch, the frames are first stretched in one direction using window muntins and then combined. Additionally, the number of horizontal and vertical window muntins in the inner frame is dynamically determined by the window’s size.

Finally, based on the position information in the rule-based layout, the adjusted components are placed in the hierarchy accordingly. The components are arranged in a structured manner, ensuring that their spatial relationships and alignment follow the intended design. This hierarchical placement ensures that the components are positioned accurately within the layout, respecting both the overall structure and the functional requirements of each individual \textcolor{revision}{sub-components}, which is highly user friendly for further \textcolor{revision}{modification}.
Furthermore, we iteratively retrieve the hierarchical structure of the rule-based layout and further enhance the alignment between components under the same node.

\label{subsec:phase2}

\section{Experments}



\begin{table}
\caption{Quantitative results of layout generation methods. \textcolor{revision}{* denotes zero-shot method.}}
    \begin{subfigure}[b]{0.45\linewidth}
        \centering
        \caption{Building layout generation.}
        \label{tab:vssota_layout}
        \begin{tabular}{@{}lrrll@{}}
        \toprule
        Method     & FID$\downarrow$   & KID$\downarrow$         &  &  \\ \midrule
        \textcolor{revision}{LayoutGPT*}  & \textcolor{revision}{94.60} & \textcolor{revision}{10.05}  &  &  \\
        ATISS      & 30.93 & 3.34            &  &  \\
        DiffuScene & 20.95 & 1.95  &  &  \\
        Ours       & 6.00  & 0.30 &  &  \\ \bottomrule

        \end{tabular}
    \end{subfigure}
    \centering
    \hfill
    \begin{subfigure}[b]{0.45\linewidth}
        \caption{Indoor layout generation.}
        \label{tab:vssota_scene_layout}
        \begin{tabular}{@{}lllll@{}}
        \toprule
        method                          & FID$\downarrow$   & KID$\downarrow$  &  &  \\ \midrule
        \textcolor{revision}{LayoutGPT*}                       & \textcolor{revision}{45.13} & \textcolor{revision}{9.89} &  &  \\
        ATISS                           & 18.60 & 1.72 &  &  \\
        DiffuScene                      & 17.21 & 0.70 &  &  \\
        Ours & 16.76 & 0.29 &  &  \\ \bottomrule
        \end{tabular}
    \end{subfigure}
    
\end{table}

In this section, we have evaluated \ourmethod and compared it with previous methods. In addition, we have demonstrated the reliability and reasonableness of the layouts generated by our method, as well as the high level of editability from our two-phase structured building generation pipeline, which were not present in earlier approaches.
\paragraph{Dataset}
The layout generation model in \ourmethod is trained on the building layout dataset we propose, the Block dataset, which contains 1.2k buildings, 42k components, and 13 component categories~(Fig.~\ref{fig:component_dist}). Furthermore, to validate the generality of our model, we follow the setup of DiffuScene~\cite{tang2024diffuscene} and conduct indoor layout generation experiments on the \emph{bedrooms} the largest subset of the 3D-Front dataset~\cite{fu20213d}. This subset contains 4,041 bedrooms and 21 component categories.

\paragraph{Baseline}
We compare our box-based layout generation method with several state-of-the-art layout generation methods, including ATISS~\cite{paschalidou2021atiss} and DiffuScene~\cite{tang2024diffuscene}. ATISS is an autoregressive model that outputs the components of a layout and their corresponding attributes in sequence. DiffuScene is a U-Net-based diffusion model that generates reasonable layouts through a step-by-step diffusion process.

\paragraph{Evaluation \textcolor{revision}{Metrics}}
To measure the realism and diversity of the generated layouts, we follow previous work and use Fréchet Inception Distance~(FID)~\cite{heusel2017gans} and Kernel Inception Distance~(KID)~\cite{binkowski2018demystifying} for evaluation. 
To apply these metrics, we represent each component of the layout with a distinct color, set transparency to 0.3, and viewed the layout from a fixed direction $(0, 0, 1)$ toward the center. The layouts are visualized as images. We generate 6,400 images for each method to ensure the reliability of the quantitative results.

To further demonstrate the effectiveness of this method, we have also conducted an additional experiment on the indoor scene task. Following the setting used in DiffuScene, we render 1,000 images through top-down orthographic projections. 
Each object category is assigned a uniform color, and textures are removed. In this case, we do not apply transparency to the colors.

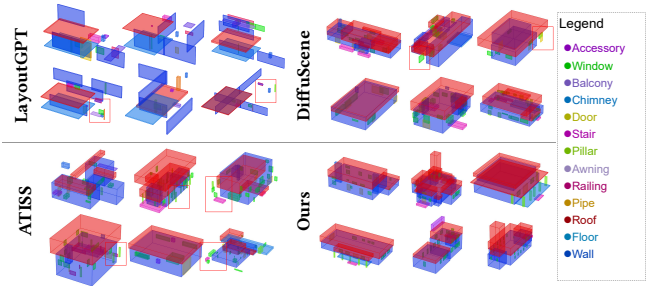
\begin{figure}[t]
    \centering
    \begin{tikzpicture}
        \node[anchor=south west, inner sep=0] (img) at (0,0) {\includegraphics[width=\linewidth]{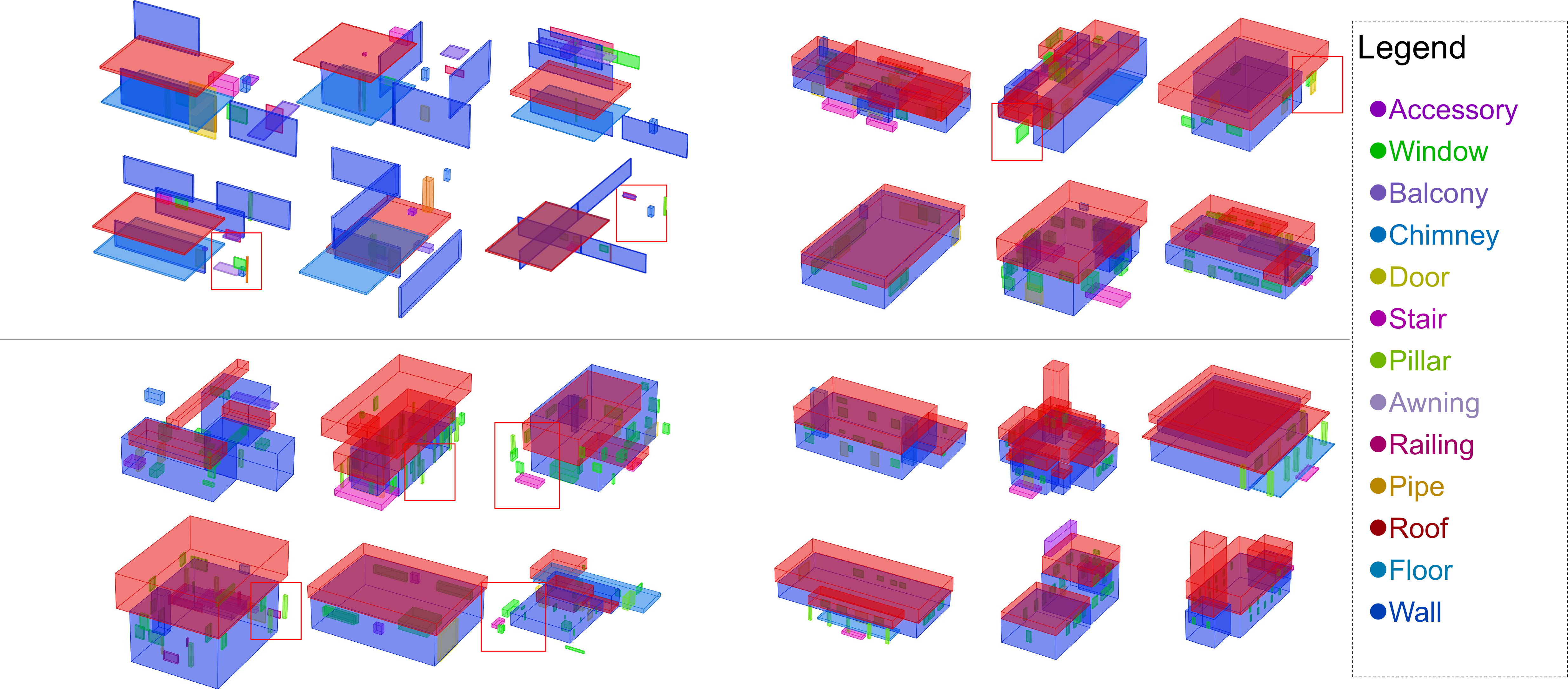}};
        
        \node[anchor=south, rotate=90, xshift=10mm, yshift=-42mm] at (img.south west) {\textbf{\footnotesize Ours}};
        \node[anchor=south, rotate=90, xshift=28mm, yshift=-42mm] at (img.south west) {\textbf{\footnotesize DiffuScene}};
        \node[anchor=south, rotate=90, xshift=10mm, yshift=-5mm] at (img.south west) {\textbf{\footnotesize ATISS}};
        \node[anchor=south, rotate=90, xshift=28mm, yshift=-5mm] at (img.south west) {\textbf{\footnotesize LayoutGPT}};
    \end{tikzpicture}
    \caption{Qualitative comparison on unconditional layout generation. It can be observed that the layouts generated by \textcolor{revision}{LayoutGPT,} ATISS and DiffuScene exhibit some issues with misaligned layouts and floating components, such as windows that are not attached to walls but instead hover in the air. Our method, on the other hand, is capable of producing high-quality layouts with coordinated components. Artifacts are marked with red boxes.}
    \label{fig:layout_cmp_demo_cropped}
\end{figure}

\paragraph{Implementation.}
We train our layout generation model on a single 3090 GPU for 50,000 epochs with batch size 64 and learning rate 2e-4. 
\textcolor{revision}{The class attribute $c$ is representated by  one-hot encoding.}

For each layout in the dataset, we apply rotation and mirroring as data augmentation techniques. 
In terms of the diffusion model, we use the default settings of the denoising diffusion probabilistic models~\cite{ho2020DDPM}, with the noise ratio ranging from 0.0001 to 0.02 over 1,000 steps. For the additional indoor layout experiments, we follow the settings used in DiffuScene~\cite{tang2024diffuscene}. The LLM used in the LGP is ChatGPT-4~\cite{achiam2023gpt}. 
Due to the presence of components embedded within walls in building generation, \ourmethod and DiffuScene do not compute IoU loss between walls and other components during training.


\subsection{Comparison with other methods}
To comprehensively evaluate the performance of \ourmethod, we compared its results with those of existing state-of-the-art generation methods. Since \ourmethod is a two-phase approach, we conducted the comparison at both the layout and architectural levels for a more thorough comparison.

\begin{figure*}
  \includegraphics[width=0.8\linewidth]{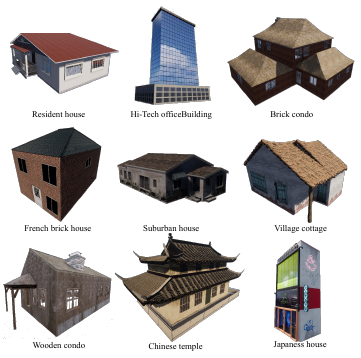}
\caption{\textcolor{revision}{
The diverse building generated with our methods across varied styles, demonstrating highly flexibility and adaptability to building style such as regions, materials, and functions. The text beneath the building indicates the style used to retrieve components during the generation process.}}
  \label{fig:add_result_3x3_cropped}
\end{figure*}

\paragraph{Box-based layout generation}
Following previous work~\cite{tang2024diffuscene}, we use two widely adopted metrics, FID and KID, to measure the distance between the distributions of ground-truth shapes and the generated ones, in order to evaluate the realism and diversity of unconditional layout generation. 
The quantitative comparison is shown in Table.~\ref{tab:vssota_layout}. 
To further validate the performance of our method, following DiffuScene, we conducted an additional experiment on the bedrooms dataset. The quantitative comparison is shown in Table.~\ref{tab:vssota_scene_layout}. Notably, for the indoor scene layout generation task, our method not only outperforms other methods comprehensively, but its generation speed is three times that of DiffuScene.
As shown in ~(Fig.\ref{fig:layout_cmp_demo_cropped}), the results generated from ATISS sometimes have disorganized layouts, with many floating components, such as windows, which only capture limited inter-component relationships, resulting in lower layout rationality. 
On the other hand, DiffuScene can capture structural information from the data. While it still has some floating components, it generates relatively reasonable simple layouts. 
In contrast, the layouts generated by our method far exceed existing box-based layout generation methods in terms of both complexity and coherence. 
\textcolor{revision}{Fig.~\ref{fig:add_result_3x3_cropped} demonstrates additional style examples to further verify the diversity and quality of results generated with our method. }

\begin{figure}
  \includegraphics[width=\linewidth]{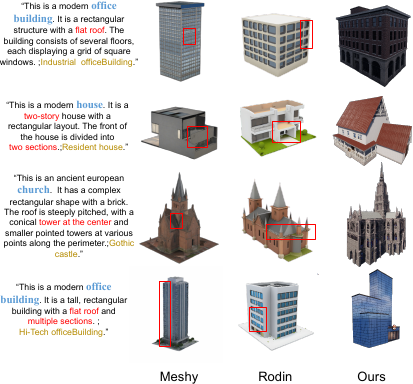}
  \caption{Qualitative comparison of buildings generated with state-of-the-art methods. Artifacts are marked with red boxes.}
  \label{fig:building_cmp_demo_cropped}
\end{figure}

\paragraph{Building generation.}
Fig.~\ref{fig:building_cmp_demo_cropped} displays the qualitative comparison, where the buildings generated by \ourmethod have a well-structured design, clear component boundaries, clean facades, and intricate details. In contrast, Meshy~\cite{meshy} tends to produce noisy surfaces and some hollow areas, while Rodin Gen-1.5~\cite{zhang2024clay} often results in overly smooth, cartoon-like surfaces with fused transitions. Our building generation pipeline far surpasses existing methods in both complexity and rationality.
\subsection{Ablation Study}
We conducted ablation experiments on the proposed approach, with results shown in Table.~\ref{tab:albation}. The results demonstrate that using real layout attributes for filling, along with encoding the physical positions of the layout, significantly improves the performance of the network. These strategies could potentially be applied to other domains involving layout generation.

\begin{table}[]
\caption{Ablation study. Using PadReal or Spatial Encoding individually only brings slight improvements, but combining them allows the model to improve realism and diversity.}
\label{tab:albation}
\begin{tabular}{ccrr}
\toprule
\textbf{PadReal} & \textbf{Spatial Encoding} & \textbf{FID$\downarrow$} & \textbf{KID$\downarrow$} \\ 
\midrule
$\times$              & $\times$                & 13.33             & 1.10             \\
$\times$               & \checkmark                        & 12.22             & 0.95              \\
\checkmark                & $\times$                    & 12.38             & 1.05              \\
\checkmark                & \checkmark                        & 6.00             & 0.30       \\
\bottomrule
\end{tabular}
\end{table}

\subsection{Editing}
\begin{figure*}
  \includegraphics[width=\linewidth]{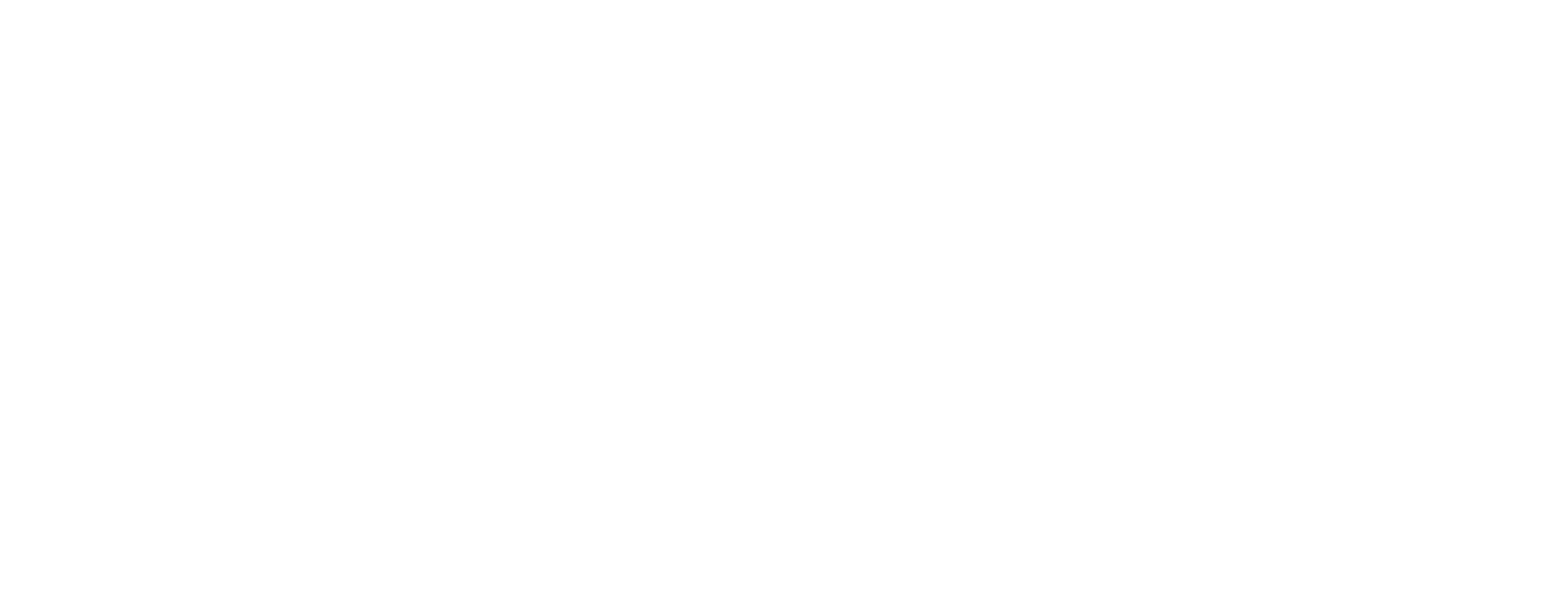}
\caption{\textcolor{revision}{
An example of progressive localized user editing. 
An initial generated building (Original) is iterative refined through progressive editing, including the addition, removal, and replacement of architectural components, as well as the modification of specific attributes such as size, material properties, and styles. Each intermediate building result includes a textual operation prompt and a hint diagram (lower-left) highlighting target layout, desired styles, or material changes for localized modifications.
}}
  \label{fig:edit}
\end{figure*}

\textcolor{revision}{Precise editing of generated results represents a fundamental user need, especially the ability to make localized adjustments to specific components without affecting others. Our two-phase approach enables precise modifications to target components through flexible adjustments while ensuring other components remain unchanged. As shown in Fig.~\ref{fig:teaser}B, we present an example of incremental editing: by gradually adding components to the layout, we achieve localized incremental edits while maintaining global stylistic consistency. Fig. \ref{fig:edit} further showcases the system's flexibility in supporting user-driven customization, where architectural components can be individually added or deleted, as long as modified with size, style, material and other component attributes, ultimately enabling a seamless transition from conceptual ideation to refined building.
}
\subsection{Limitations and Future Works}

\begin{figure*}
    \centering
    \begin{minipage}{0.49\textwidth}
        \centering
        \begin{tikzpicture}
        \node[anchor=center] (img) { 
            \includegraphics[width=1.0\linewidth]{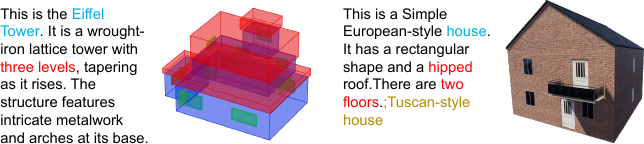}
        };
              \node[anchor=south west, xshift=20mm, yshift=-5mm] at (img.south west) 
            {(a)};
              \node[anchor=south west, xshift=-25mm, yshift=-5mm] at (img.south east) 
            {(b)};
        \end{tikzpicture}
    \caption{\textcolor{revision}{Failure cases. a) An example where the LGP fails to generate rare building layouts due to lack of similar data in the training dataset.b) An example where the generated building do not fully align with the intended style when the specified style exceeds the supported range, defaulting to the closest available style. }}
      \label{fig:failure_case}
    \end{minipage}%
    \hfill
    \begin{minipage}{0.49\textwidth}
            \centering
            \begin{subfigure}[b]{0.4\linewidth}
                \centering
                \includegraphics[width=\linewidth]{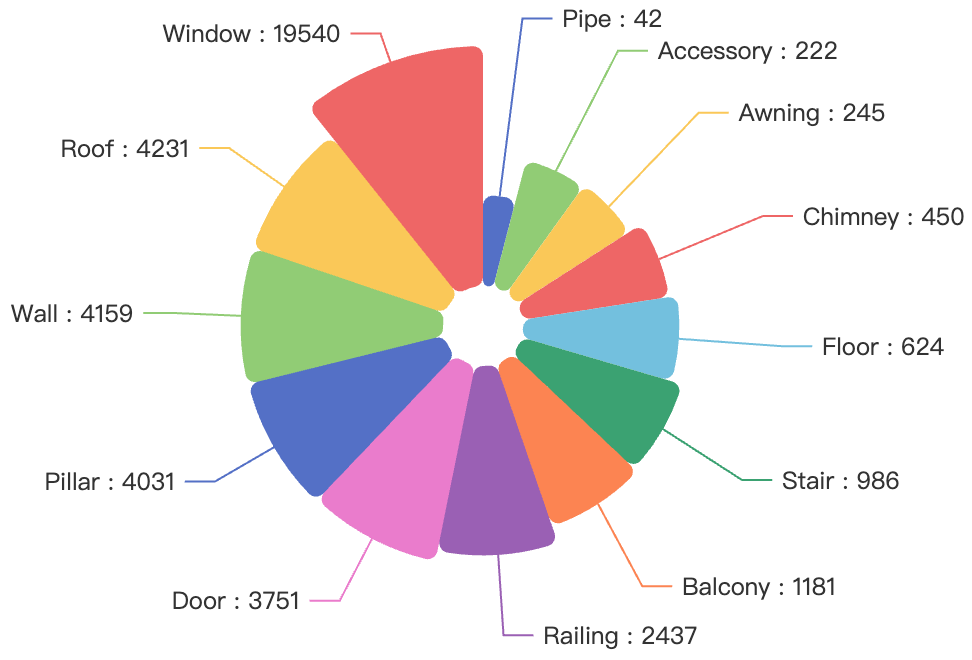}
            \caption{Box label.}
            \Description{}
            \label{fig:component_dist}
            \end{subfigure}
            \centering
            \hspace{5pt}
            \begin{subfigure}[b]{0.4\linewidth}
                \centering
                \includegraphics[width=\linewidth]{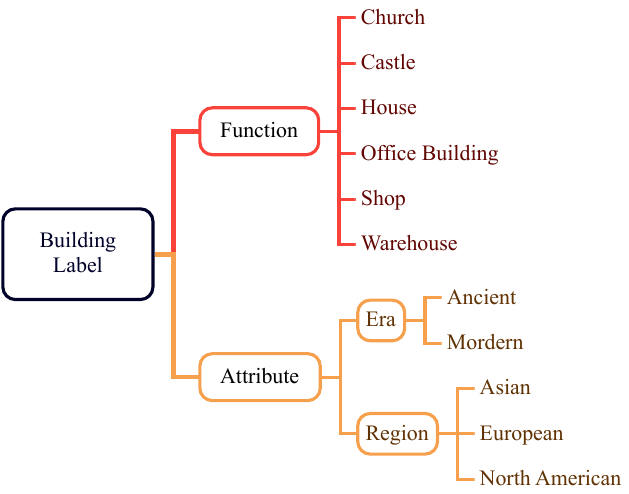}
                \caption{Building label.}
                \Description{}
                \label{fig:style_label}
            \end{subfigure}
        
        \caption{Labels of building in our dataset. 
        a) The distribution of instances across each box category in the dataset. We visualized this using a Nightingale's Rose Diagram~\cite{brasseur2005florence} with a logarithmic scale for better clarity. 
        b) A diagram of building label types. Buildings are first categorized by function, and then additional attributes are assigned based on regional and era styles.
        }
    \end{minipage}
\end{figure*}

\textcolor{revision}{
LGP may fails to perfectly generate layout with significantly divergent structures that are absent from the training set, such as the Eiffel Tower in Fig.~\ref{fig:failure_case} (a).
In addition, due to the finite scope of the component asset library, when encountering unsupported building styles, the LLM-based layout will attach components with the closest available style type, potentially resulting in a misalignment between the generated building's style and the desired one (Fig.~\ref{fig:failure_case} (b)).
To address these two issues, we propose two key directions for future work:
1) Develop automated building annotation methods based on existing data to improve LGP's generalization across unseen architectural forms.
2) Integrate a variety of generative models into our methods to create assets, textures, decals, etc., to dynamically expand our database and automate the extension of new styles.
}



\section{Conclusion}

In this work, we present \ourmethod, a hybrid approach for text-driven structured buildings generation. In the Layout Generation Phase, a Transformer-based diffusion model is employed to produce box-based layouts, which are completed by a pre-trained LLM to address semantic gaps and extend them into rule-based layouts. 
These rule-based layouts are then inputted into the Building Construction Phase to generate structured buildings through PCG. To support this pipeline, we introduce the first 3D architectural layout dataset, comprising box-based layouts paired with the corresponding textual descriptions. Our method surpasses state-of-the-art approaches in layout metrics and visual details, while maintaining global consistency and hierarchical structure. 
Moreover, \ourmethod is highly editing-friendly, enabling localized modifications to bounding boxes for refining specific parts of the generated building while maintaining high quality, global consistency, and a hierarchical structure.
This approach establishes a robust foundation for generating controllable, structured 3D buildings, advancing both the methodology and dataset resources in the field.

\begin{acks}
We sincerely thank the anonymous reviewers for their professional, insightful, and constructive comments, which have helped us improve the quality and clarity of this paper. Weiwei Xu is partially supported by NSFC~(No. 62421003). This paper is supported by the Information Technology Center and State Key Lab of CAD\&CG, Zhejiang University.
\end{acks}
\bibliographystyle{ACM-Reference-Format}
\bibliography{reference}

\newpage

\clearpage

\end{document}